# A Comprehensive Study on the Molecular Dynamics of Pristine and Functionalized Graphene Nanoplatelets

*Olasunbo Z. Farinre[†], Hawazin Alghamdi[†], Matthew L. Kelley[§], Adam J. Biacchi[§], Albert V. Davydov[£], Christina A. Hacker[§], Albert F. Rigosi[§], and Prabhakar Misra[†*]*

[†]Department of Physics and Astronomy, Howard University, Washington, DC 20059, USA

[§]Physical Measurement Laboratory, National Institute of Standards and Technology, Gaithersburg, MD 20899, USA

[£]Material Measurement Laboratory, National Institute of Standards and Technology, Gaithersburg, MD 20899, USA

[*]Corresponding Author: Prabhakar Misra; pmisra@howard.edu; Tel. +1 (202) 806-6251

ABSTRACT Graphene nanoplatelets (GnPs) are promising candidates for gas sensing applications because they have a high surface area to volume ratio, high conductivity, and a high temperature stability. Also, they cost less to synthesize, and they are lightweight, making them even more attractive than other 2D carbon-based materials. In this paper, the surface and structural properties of pristine and functionalized GnPs, specifically with carboxyl, ammonia, carboxyl, nitrogen, oxygen, fluorocarbon, and argon, were examined with Raman spectroscopy, Fourier transform infrared spectroscopy, X-ray photoelectron spectroscopy and X-ray diffraction (XRD) to determine the functional groups present and effects of those groups on the structural and vibrational properties. We attribute certain features in the observed Raman spectra to the variations in concentration of the functionalized GnPs. XRD results show smaller crystallite sizes for functionalized GnPs samples that agree with images acquired with scanning electron microscopy. Lastly, a molecular dynamics simulation is employed to gain a better understanding of the Raman and adsorption properties of pristine GnPs.

1. INTRODUCTION

The growing global concern for environmental pollution due to its negative impact on the Earth's climate system and human health has led to an increased need for gas sensors with useful properties such as a high sensitivity and selectivity. These properties are important for the identification, monitoring and removal of toxic gases from the environment. Graphene and its derivatives such as graphene oxide, reduced graphene oxide, and carbon nanotubes (single-walled and multi-walled CNTs) have been extensively researched for gas sensing applications due to their high surface area[1][2][3][4]. The selective chemical functionalization of graphene makes them attractive for a variety of high-impact applications such as aerospace polymer reinforcement [5,6], electrical metrology [7-9], and most relevantly to this work, gas sensing applications, especially when compared to commonly used metal oxide semiconductors. Overall, metal oxide semiconductors can be good gas sensing materials due to their low cost, high sensitivity, and ease-of-fabrication; however, some of their drawbacks are poor selectivity, short life span, and the need for higher temperatures for their operability[10].

Graphene nanoplatelets (GnPs) have become an alternative to graphene because of their exciting properties such as excellent thermal and electrical conductivities, high mechanical rigidity, high aspect ratio, light weight, and the potential for large-scale production compared to graphene. GnP sheets often aggregate into flakes consisting of weakly interacting monolayer sheets due to strong Van der Waals attractions and strong hydrophobicity, and their functionalization leads to enhancement of the adsorption of target molecules on its basal plane. GnPs have the same honeycomb structure as graphene, wherein the carbon atoms are arranged in a hexagonal ring with an atomic distance of 1.42 Å.[#] Each carbon atom is connected to three other neighboring carbon atoms via three $\sigma$-bonds and one $\pi$-bond. Recently, research studies have reported the addition of GnPs to other materials (*e.g.* metal oxides and polymers) to form nanocomposites in order to enhance their gas sensing abilities[11-13]. This paper focuses on the morphology, surface, and structural properties of pristine and functionalized GnPs.

In this work, commercially acquired pristine and functionalized GnPs were examined using Raman spectroscopy, X-ray photoelectron spectroscopy (XPS), X-ray diffraction (XRD), and scanning electron microscopy (SEM) to identify the functional groups present in the variety of GnP samples and to determine the effects of those functional groups on surface properties. Data acquired with Raman spectroscopy suggest the introduction of *n*-type doping when **35 wt.%** carboxyl functional groups are added to pristine GnPs. Furthermore, the effects of the same functional groups on structural properties, as well as molecular dynamics simulations, are discussed.

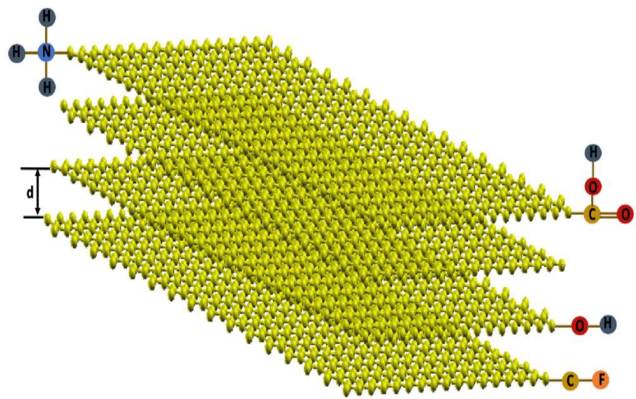

**Figure 1.** Diagram showing functional groups (ammonia, fluorocarbon, hydroxyl, and carboxyl) attached to the edges of one graphene nanoplatelet sheet consisting of approximately four layers of graphene. The interlayer distance (*d*) is 3.35 Å.

## 2. MATERIALS AND METHODS

### 2.1. Experimental

*2.1.1. Pristine and Functionalized Graphene Nanoplatelets (GnPs)*

The pristine GnPs (GnPs-P) were acquired from US Research Nanomaterials, Inc. (see notes), whereas GnPs functionalized with low-density ammonia (GnPs-NH$_3$), carboxyl (GnPs-COOH), argon (GnPs-A), nitrogen (GnPs-N), oxygen (GnPs-O), and fluorocarbon (GnPs-CF) functional groups were acquired from Graphene Supermarket (see notes). The 35 wt.% carboxyl functionalized GnPs (GnPs-35COOH) was acquired from Cheap Tubes (see notes). No chemical alterations or modifications were imposed on the acquired samples.

*2.1.2. Characterization of GnP Samples*

Raman spectroscopy data were recorded on Renishaw inVia Raman spectrometer using a

514 nm wavelength excitation laser source, laser beam quality of 0.65 mm and laser maximum power of 50 mW (see notes). The spectra were collected using a laser exposure time of 10 s with 10 accumulations to reduce the signal-to-noise ratio.

The XRD spectra of GnPs-P, GnPs-NH$_3$, GnPs-COOH, GnPs-A, GnPs-N, GnPs-O and GnPs-CF were collected using the Thermo Scientific ARL$^{TM}$ EQUINOX 100 X-Ray diffractometer (see notes) with a wavelength of 1.54056 Å whereas the XRD spectrum of GnPs-35COOH was collected using the Philips X'Pert powder X-ray Diffractometer (see notes) with Cu K$_\alpha$ radiation. XPS measurements were performed using the Kratos X-ray Photoelectron Spectrometer-Axis Ultra DLD (see notes) operating at a base pressure of 2.0 x 10$^{-9}$ Torr. All samples were analyzed using monochromatic Al K-alpha source (1486.7 eV) with spot size 300 μm x 700 μm. Pass energies of 160 eV and 20 eV were used to collect the survey and high-energy resolution core level XPS spectra, respectively.

## 2.2. Molecular Dynamics Simulation

Investigating the phonon properties requires calculating the dynamical matrix directly from molecular dynamics (MD) simulations. The dynamical matrix (D) given in equation [1] is constructed by evaluating the displacements of atoms during the MD simulation using fluctuation-dissipation theory, while taking into account the anharmonicity of phonons in the model system[14]. This method is implemented by a "Fix Phonon" command incorporated within the Large-scale Atomic /Molecular Massively Parallel Simulator (LAMMPS) software package from Sandia National Laboratories[15]. The computed dynamical matrices are then passed to a post-processing code to evaluate the vibrational properties.

$$D_{k\alpha,k'\beta}(q) = (m_k m_{k'})^{-1/2} \, \phi_{k\alpha,k'\beta}(q) \tag{1}$$

Where $\phi_{k\alpha,k'\beta}(q)$ is the force constant coefficient of the system in reciprocal space.

LAMMPS is a MD simulation software code that can run efficiently and effectively on parallel computers. Also, it can model micro-canonical (NVE), canonical (NVT) and grand-canonical ensembles of a system of particles in liquid, solid, or gaseous state using a variety of fields, potentials, and boundary conditions[16].

## 3. RESULTS AND DISCUSSION

### 3.1. Morphology of Pristine and Functionalized GnPs

The scanning electron microscope (SEM) gives information about the morphology of all GnP species. The Phenom Pure SEM (see notes), with magnification that ranges from 20× to 65000×, was utilized to obtain the SEM images with 3D images being generated with Gwyddion[17] as shown in Fig. 2. The SEM images of GnPs-P show that the platelets are randomly stacked on each other forming aggregates, a behavior that suggests a strong hydrophobicity and Van der Waals interaction. Individual nanoplatelet sheets consist of 3-6 layers of graphene.

It was observed that, when compared with GnPs-P, the functionalized GnPs were generally fragmented and of smaller lateral sizes, as shown in table 1. This can be attributed to the dielectric barrier discharge plasma method employed in the synthesis of the functionalized samples[18]. For instance, during graphene oxide (GO) synthesis, research has shown that acids and oxidizing agents can break the GO sheets into smaller lateral sizes when exposed for a long oxidation time[19].

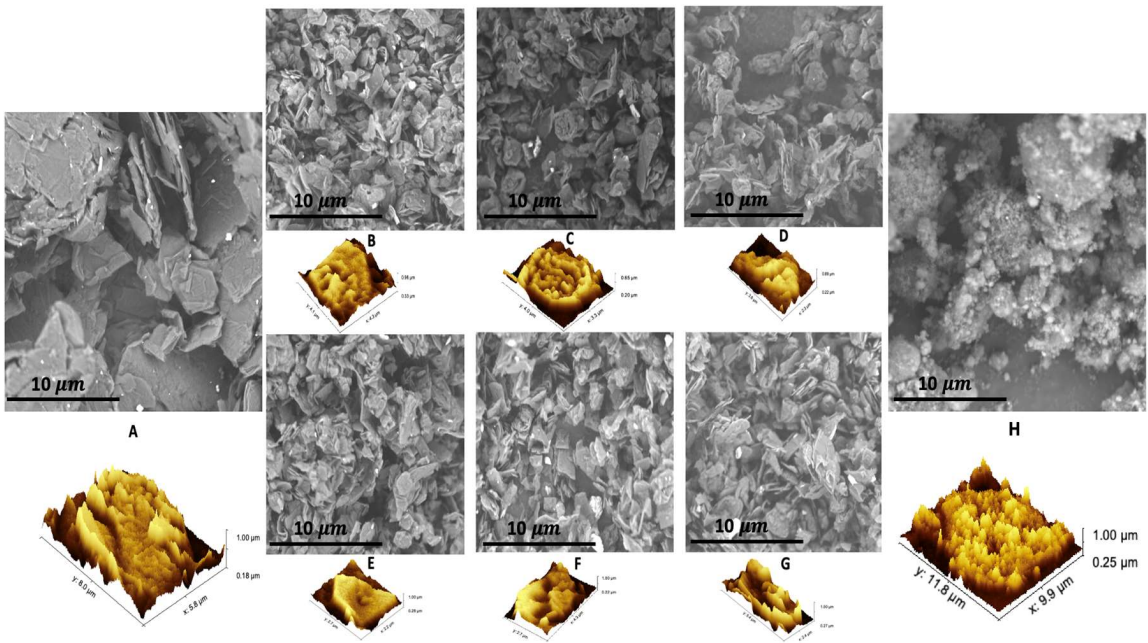

**Figure 2.** SEM and 3D images of (A) GnPs-P (B) GnPs-NH$_3$ (C) GnPs-A (D) GnPs-COOH (E) GnPs-CF (F) GnPs-N (G) GnPs-O (H) GnPs-35COOH taken at 10 $\mu$m resolution.

The SEM image of the GnPs-35COOH shows that the platelets become smaller to form conglomerates because the edges become delaminated[20] and the presence of a high concentration of oxygen-containing functional groups in the sample. Table 1 shows the lateral dimensions and thicknesses of the platelets' aggregates within pristine and functionalized GnPs depending on the number platelets stacked in the aggregates. The dimensions were calculated from the average of four aggregates cropped from different spots within the SEM images.

**Table 1.** Lateral dimensions and thicknesses of pristine and functionalized GnPs' aggregates

| GnPs Samples | X Average ($\mu m$) | Y Average ($\mu m$) | Z Average ($\mu m$) |
|---|---|---|---|
| GnPs-P | 5.20 | 6.40 | 0.79 |
| GnPs-NH$_3$ | 3.60 | 3.50 | 0.70 |
| GnPs-A | 3.30 | 3.40 | 0.65 |
| GnPs-COOH | 3.10 | 4.50 | 0.70 |
| GnPs-35COOH | 10.0 | 10.2 | 0.73 |
| GnPs-CF | 3.0 | 2.90 | 0.69 |
| GnPs-N | 3.90 | 2.80 | 0.76 |
| GnPs-O | 2.90 | 3.80 | 0.65 |

### 3.2. Raman spectra of Pristine and Functionalized GnPs

The vibrational properties of pristine and functionalized GnPs were investigated by Raman spectroscopy. Three major peaks were observed in the Raman spectra of pristine and functionalized GnPs samples shown in Fig. 3, namely the D peak (between 1349 cm$^{-1}$ and 1357 cm$^{-1}$), which indicates the presence of disorder in the sp$^2$ carbon lattice[21], the G peak (between 1580 cm$^{-1}$ and 1583 cm$^{-1}$), attributed to the in-plane stretching vibration of the sp$^2$ bonded carbon atoms[22], the 2D peak (between 2707 cm$^{-1}$ and 2723 cm$^{-1}$), which originates from the combination of two Raman A$_{1g}$ modes and lastly, the 2D' peak, representing a second order mode of the D' peak[23]. The Raman spectra of GnPs-COOH, GnPs-35COOH, GnPs-A, GnPs-NH$_3$, GnPs-N, GnPs-O and GnPs-CF show an increase in the D band intensities compared to GnPs-P indicating the formation of defects in the structures. In addition, the observed D' peak in all functionalized GnPs samples' spectra may be attributed to the presence of the functional groups introduced and is activated by two-phonon double resonance Raman scattering involving one longitudinal optical (LO) phonon near the gamma ($\Gamma$) point[24].

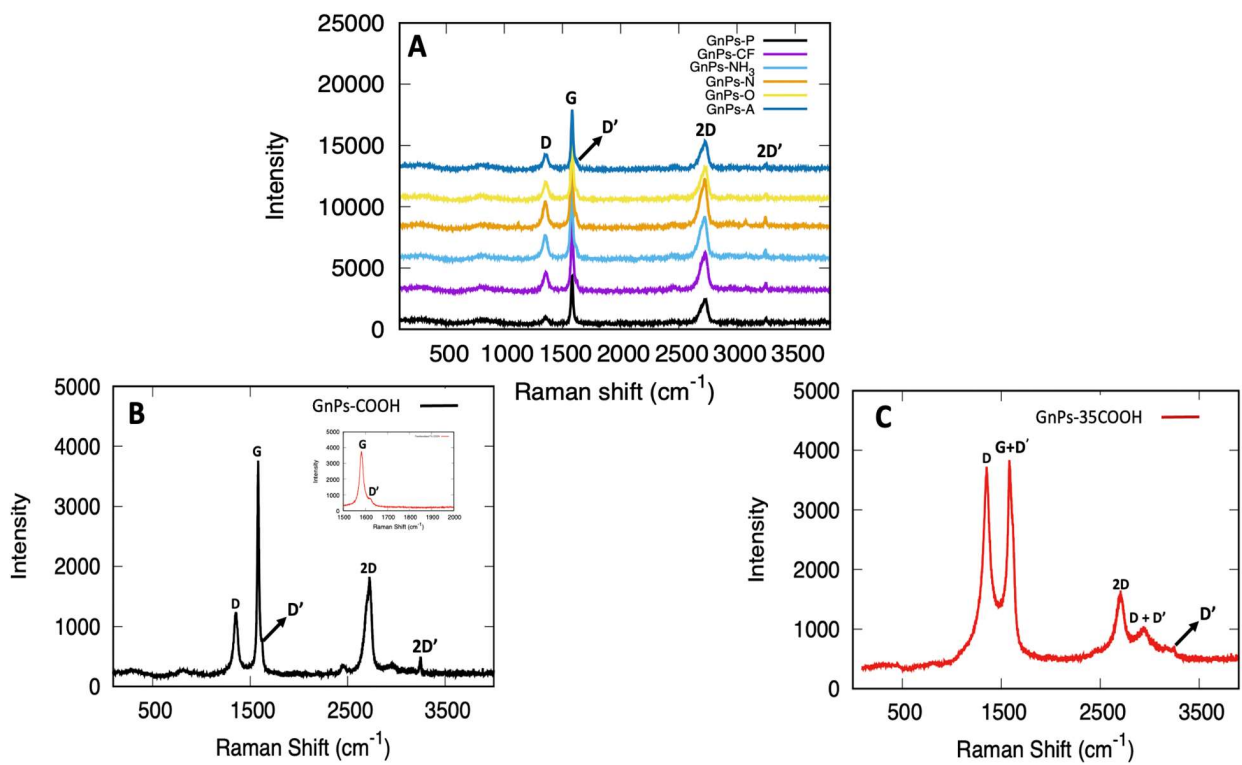

**Figure 3.** Raman spectra of (A) GnPs-P, GnPs-NH$_3$, GnPs-N, GnPs-O, GnPs-CF and GnPs-A (B) GnPs-COOH (C) GnPs-35COOH.

However, the D' peak merges with the G peak forming the G+D' in the Raman spectrum of GnPs-35COOH as shown in Fig. 3(C). This is because of the G-band peak broadening indicating a higher degree of disorder in the structure compared to the low density functionalized GnPs samples. The Raman spectrum of GnPs-35COOH is characterized by intense D and broad 2D peaks. The increase in intensity of the D peak of GnPs-35COOH in relation to the D peak of the low density-functionalized GnPs indicates an increase in the amount of the disordered phase. Also, the appearance of the D+D' peak in the GnPs-35COOH spectrum confirms the formation of structural defects due to an increase in the concentration of functional groups present.

Table 2 shows the average intensity ratios of D and G ($I_D/I_G$), 2D and G ($I_{2D}/I_G$) peaks and Raman positions of the D, G and 2D peaks for pristine and functionalized GnPs. A consequence of doping with the functional groups is a significant increase in the average value of $I_D/I_G$ with respect to the concentration of doping introduced. This explains the reason GnPs-35COOH has the highest $I_D/I_G$ average value. As shown in table 2, the $I_D/I_G$ average values of the low density functionalized GnPs do not change much but a significant decrease is observed in the $I_D/I_G$ average of GnPs-35COOH. This result shows that the carboxyl group introduces electron doping to pristine GnPs[25][26]. A significant Raman shift is not observed in the G peak frequency of all the functionalized GnPs samples indicating the functional groups introduced are not causing a significant amount of strain within the structure. Also, table 2 shows that there is a red-shift in the 2D peak frequency of GnPs-35COOH which confirms the introduction of electron doping when carboxyl is introduced. The frequency of the 2D band is expected to shift to lower frequencies upon *n*-type doping in graphene based on earlier research studies[27][28]. Also, a slight decrease in the $I_{2D}/I_G$ value of GnPs-35COOH is observed which confirms the *n*-type behavior of GnPs-35COOH[25].

**Table 2.** Intensity ratio values and position of the bands in the Raman spectra of pristine and functionalized GnPs

| GnPs Samples | Band D (cm$^{-1}$) | Band G (cm$^{-1}$) | Band 2D (cm$^{-1}$) | $I_D/I_G$ | $I_{2D}/I_G$ |
|---|---|---|---|---|---|
| GnPs-P | 1354± 3.3 | 1581 ± 1.1 | 2722 ± 2.5 | 0.25 ± 0.10 | 0.58 ± 0.10 |
| GnPs-NH$_3$ | 1355± 2.2 | 1582 ± 1.4 | 2722 ± 2.3 | 0.36 ± 0.01 | 0.51 ± 0.05 |
| GnPs-A | 1356± 2.2 | 1583 ± 0.2 | 2723 ± 1.4 | 0.34 ± 0.01 | 0.53 ± 0.03 |
| GnPs-COOH | 1354± 2.0 | 1581 ± 1.7 | 2721 ± 3.7 | 0.33 ± 0.03 | 0.52 ± 0.03 |
| GnPs-35COOH | 1350± 5.4 | 1581 ± 1.8 | 2708 ± 5.4 | 0.93 ± 0.04 | 0.41 ± 0.04 |
| GnPs-CF | 1356± 1.4 | 1582 ± 0.7 | 2723 ± 3.3 | 0.29 ± 0.02 | 0.50 ± 0.07 |
| GnPs-N | 1355± 1.0 | 1581 ± 0.9 | 2720 ± 1.0 | 0.36 ± 0.03 | 0.50 ± 0.02 |
| GnPs-O | 1357± 0.6 | 1583 ± 0.3 | 2724 ± 1.7 | 0.32 ± 0.01 | 0.52 ± 0.02 |

**Table 3.** Functional groups present in pristine and functionalized GnPs

| GnPs-Samples | Functional Groups | Wavenumber (cm$^{-1}$) | GnPs-Samples | Functional Groups | Wavenumber (cm$^{-1}$) |
|---|---|---|---|---|---|
| GnPs-P | Free O-H<br>C-H (bend)<br>C=C<br>O-H (bend)<br>C-O<br>C-H | 3735.00<br>2685.00<br>1670.00<br>1431.00<br>1291.60<br>872.420 | GnPs-N | N-H (stretch)<br>C-O (stretch) | 2969<br>1054 |
| GnPs-COOH | C=O<br>C-O | 1597.88<br>1148.90 | GnPs-CF | C-O | 1036 |
| GnPs-35COOH | Free O-H<br>C-H (stretch)<br>C=O (symmetric stretch)<br>C=O (asymmetric stretch)<br>O-H (bend)<br>C-O | 3749.00<br>2658.00<br>1687.00<br>1844.00<br>1539.00<br>992.00 | GnPs-A | C-O | 1036 |
| GnPs-NH$_3$ | N-H (stretch)<br>C-O (stretch) | 2639<br>1054 | | | |
| GnPs-O | C=O<br>C-O | 1608<br>1054 | | | |

## 3.3. X-ray Photoelectron spectra of Pristine and Functionalized GnPs

The surface composition of the pristine and functionalized GnPs was analyzed using XPS. Figs. 5, 6, 7 and 8 give the high-resolution C 1s, N 1s, O 1s and F 1s spectra of GnPs-P, GnPs-COOH, GnPs-35COOH, GnPs-A, GnPs-NH$_3$, GnPs-N, GnPs-O and GnPs-CF. Table 4 reports the individual contributions within the high-resolution spectra of the GnPs materials. The deconvoluted C 1s spectra of GnPs-P is characterized by contributions at 284.5 eV, 285.2 eV and 287.2 eV arising from C=C (sp$^2$ bonded carbon atoms), C-C (sp$^3$ bonded carbon atoms), C-O (carbonyl), respectively, whereas the deconvoluted N and O 1s spectra show the C-N and C-O moieties, respectively. The presence of the C=C peak in the XPS spectra of GnPs-COOH, GnPs-35COOH, GnPs-A, GnPs-NH$_3$, GnPs-N, GnPs-O and GnPs-CF samples shows that the functional groups do not disrupt the sp$^2$ graphitic carbon structure.

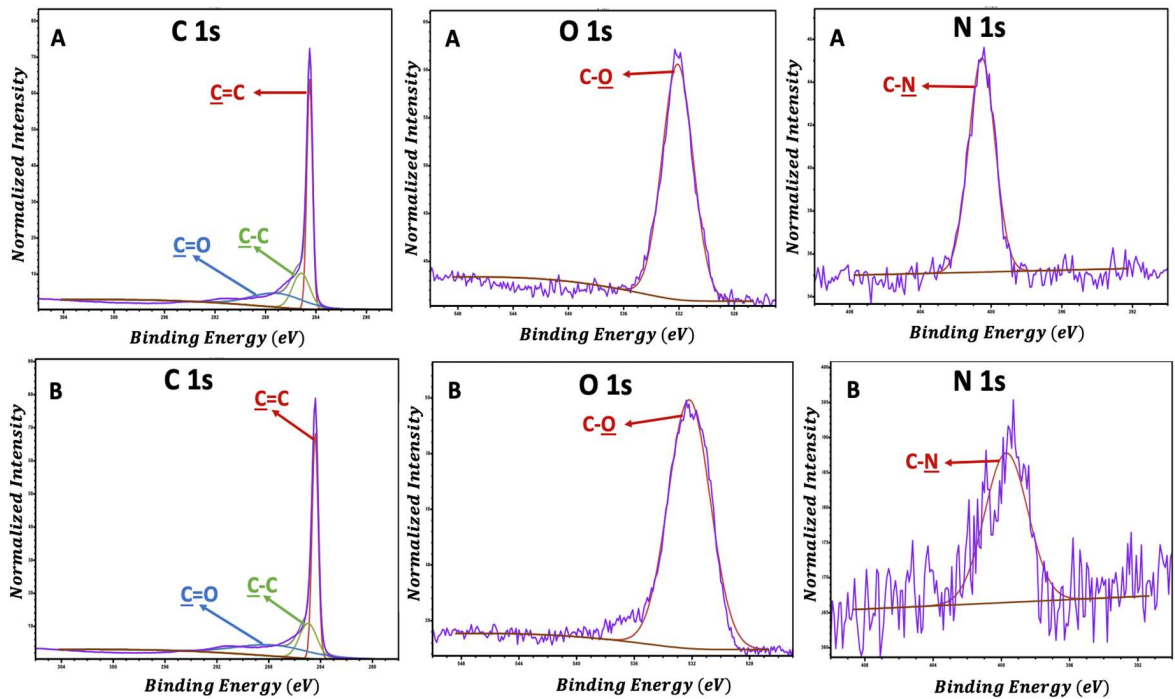

**Figure 5.** High-resolution core-level C 1s, N 1s and O 1s XPS spectra of (A) GnPs-P and (B) GnPs-NH$_3$

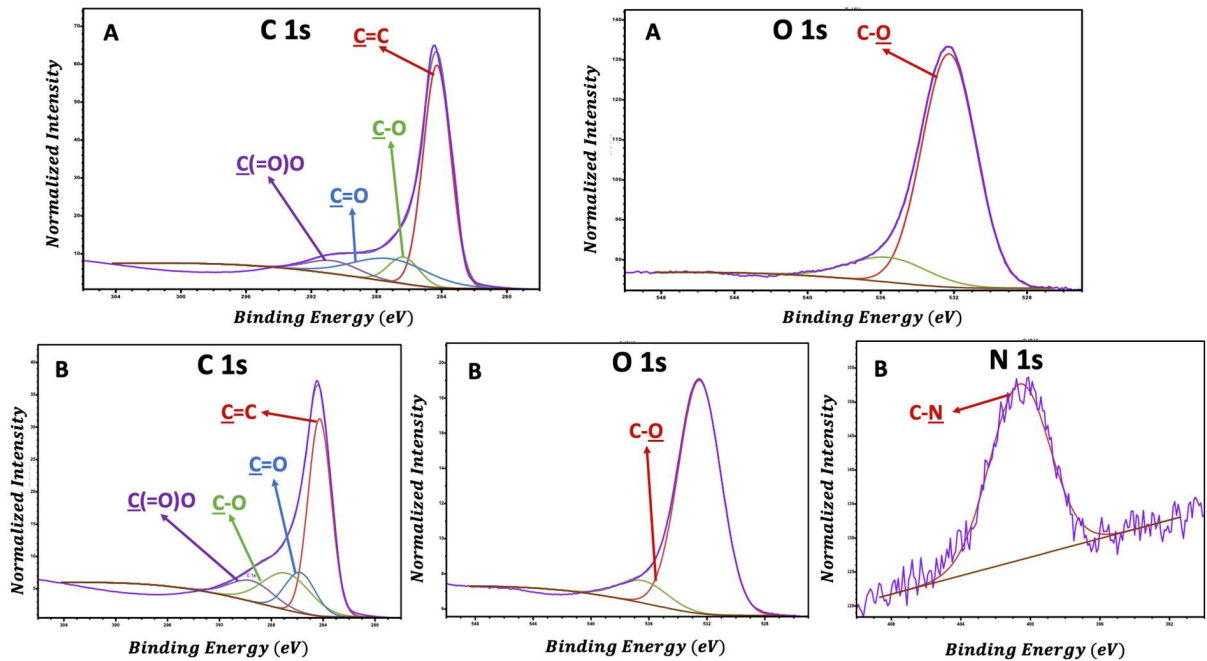

**Figure 6.** High-resolution core-level (A) C 1s and O 1s XPS spectra of GnPs-COOH (B) C 1s, O 1s and N 1s XPS spectra of GnPs-35COOH

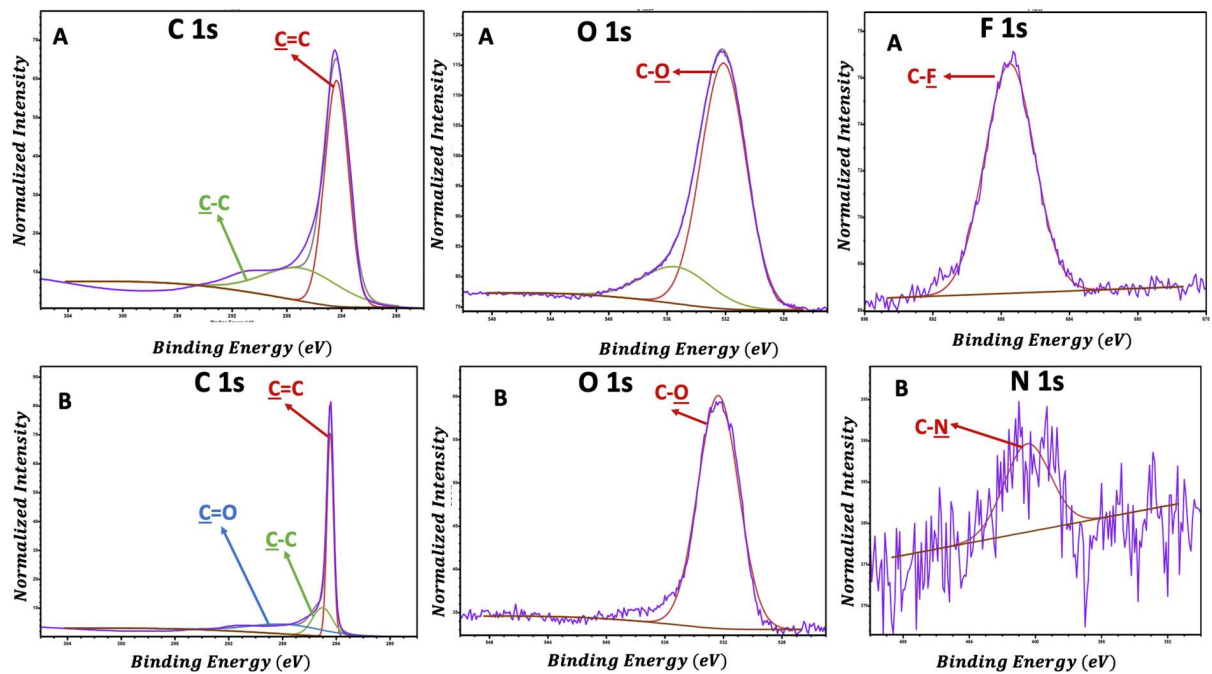

**Figure 7.** High-resolution core-level (A) C 1s, O 1s and F 1s XPS spectra of GnPs-CF (B) C 1s, O 1s and N 1s XPS spectra of GnPs-N

**Figure 8.** High-resolution core-level C 1s and O 1s XPS spectra of (A) GnPs-A and (B) GnPs-O

**Table 4.** Contributions of individual chemical moieties in the high-resolution C 1s, N 1s, O 1s and F 1s spectra of GnPs-P, GnPs-COOH, GnPs-35COOH, GnPs-NH$_3$, GnPs-O, GnPs-N, GnPs-A and GnPs-CF

|  | Functional Groups | GnPs-P (eV) | GnPs-COOH (eV) | GnPs-35COOH (eV) | GnPs-NH$_3$ (eV) | GnPs-O (eV) | GnPs-N (eV) | GnPs-A (eV) | GnPs-CF (eV) |
|---|---|---|---|---|---|---|---|---|---|
| C 1s | C=C<br>C-C<br>C-O<br>C=O<br>C(=O)O | 284.5<br>285.2<br>-<br>287.2<br>- | 284.3<br>-<br>286.2<br>287<br>290.6 | 284.3<br>-<br>285.8<br>286.8<br>289.5 | 284.4<br>285.0<br>-<br>287.7<br>- | 284.5<br>285.0<br>-<br>287.7<br>- | 284.4<br>285.0<br>-<br>287.7<br>- | 284.5<br>285.0<br>-<br>288.0<br>- | 284.4<br>-<br>-<br>286.9<br>- |
| N 1s | C-N | 400.5 | - | 400.6 | 399.7 | - | 401 | - | - |
| O 1s | C-O | 532.1 | 532.4 | 532.5 | 532.2 | 532.4 | 532.4 | 532.4 | 532.1 |
| F 1s | C-F | - | - | - | - | - | - | - | 687.5 |

**Table 5.** Elemental surface composition of GnPs-P, GnPs-COOH, GnPs-35COOH, GnPs-NH$_3$, GnPs-O, GnPs-N, GnPs-A and GnPs-CF

|      | GnPs-P (at.%) | GnPs-COOH (at.%) | GnPs-35COOH (at.%) | GnPs-NH$_3$ (at.%) | GnPs-O (at.%) | GnPs-N (at.%) | GnPs-A (at.%) | GnPs-CF (at.%) |
|------|---------------|------------------|--------------------|--------------------|---------------|---------------|---------------|----------------|
| C 1s | 96.94 | 96.20 | 88.32 | 97.00 | 95.30 | 96.74 | 97.39 | 97.83 |
| N 1s | 1.16 | - | 0.33 | 0.25 | - | 0.19 | - | - |
| O 1s | 1.90 | 3.80 | 11.35 | 2.75 | 4.00 | 3.01 | 2.60 | 2.75 |
| F 1s | - | - | - | - | - | - | - | 0.39 |

Table 5 summarizes the surface composition of elements, as determined by XPS: carbon (C), nitrogen (N), oxygen (O) and fluorine (F). Results show a significant increase and decrease of oxygen and carbon elemental composition, respectively, in the spectrum of GnPs-35COOH when compared with GnPs-P, GnPs-COOH, GnPs-A, GnPs-N, GnPs-O, GnPs-CF and GnPs-NH$_3$. In addition, low elemental composition of nitrogen in GnPs-NH$_3$ and GnPs-N, oxygen in GnPs-COOH and GnPs-O and lastly, fluorine in GnPs-CF was observed are recorded in table 5.

### 3.4. X-ray Diffraction spectra of Pristine and Functionalized GnPs

The structural quality of pristine and functionalized GnPs was determined by XRD. The (002), (100), (004), (221) and (110) diffraction peaks were observed in the XRD spectra of GnPs-P, GnPs-COOH, GnPs-A, GnPs-NH$_3$, GnPs-N, GnPs-O and GnPs-CF shown in Fig. 9 where each of the peaks represent the planes in the hexagonal crystal lattice of GnPs. The high intensity diffraction peak found at 38° is from the powder sample holder in the Thermo Scientific X-ray diffractometer utilized in collecting the spectra. The high intensity (002) diffraction peak found between 26.36° and 26.71° in the XRD spectra of GnPs-P, GnPs-COOH, GnPs-A, GnPs-NH$_3$, GnPs-O, GnPs-N and GnPs-CF confirms a high degree of crystallinity of the samples[30]. The (002) diffraction peaks of the pristine and functionalized GnPs samples correspond to an interlayer spacing ($d_{002}$) of 0.34 nm which was calculated using Bragg's formula in equation [2]; the value was found to agree well with the $d_{002}$ value of graphite[31]. This result suggests that the functional groups present do not cause significant strain within the structure or lattice expansion.

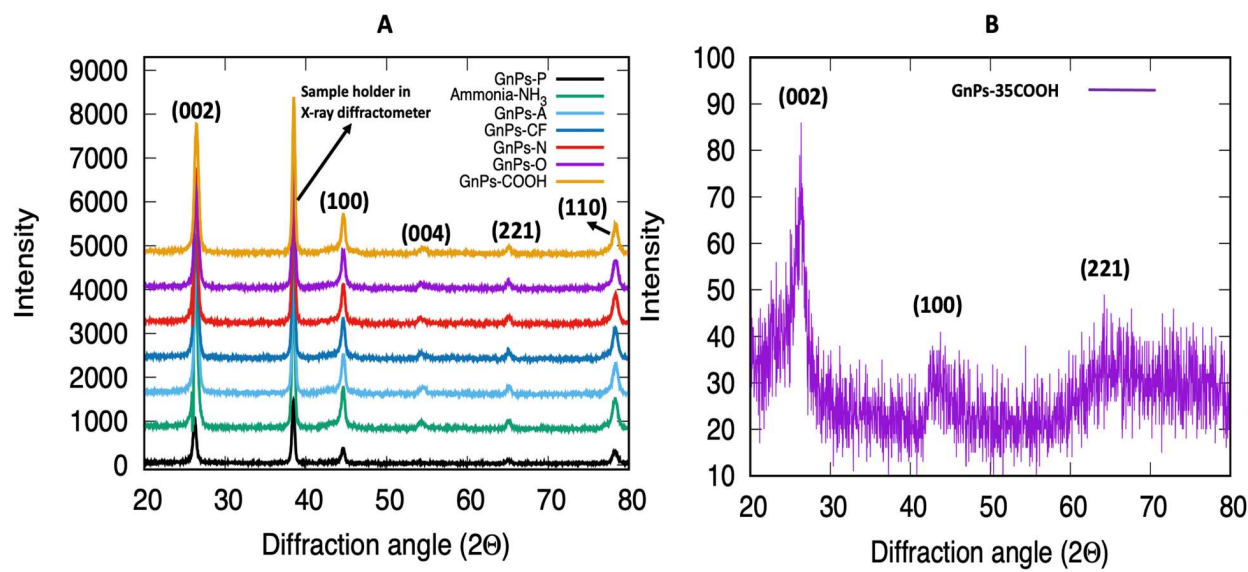

**Figure 9.** XRD spectra of (A) GnPs-P, GnPs-NH$_3$, GnPs-O, GnPs-COOH, GnPs-N, GnPs-CF, and GnPs-A (B) GnPs-35COOH

However, the XRD spectra of GnPs-35COOH show lower intensity and broader peaks indicating a low structural quality and highly disordered GnP sheets. The in-plane ($L_a$) and out-of-plane distances along the *c*-direction ($L_c$), and other relevant length scales are shown in table 5. Some values were calculated from the full width half maximum (FWHM) values of the (100) and (002) peaks using the Scherrer equation (equations [3] and [4])[32]. Results shows that the crystallite sizes of the functionalized GnPs samples became smaller which agrees with our SEM results explained earlier. Also, GnPs-35COOH was observed to have a smaller crystallite size compared to the GnPs-COOH, GnPs-A, GnPs-NH3, GnPs-O, GnPs-N and GnPs-CF samples suggesting a distortion or delamination of the edges to a large extent. Table 5 also displays the overall number of layers along the *c*-direction ($N_c$)[33] in GnPs' aggregates, calculated using equation [5]. The results show that GnPs-35COOH have the lowest number of layers demonstrating a high degree of exfoliation from graphite.

$$d_{002} = \frac{n\lambda}{sin\theta} \qquad (2)$$

$$L_c = \frac{0.89\lambda}{FWHM\ (002)(2\theta)\ X\ cos\theta} \qquad (3)$$

$$L_a = \frac{0.89\lambda}{FWHM\ (100)(2\theta)\ X\ cos\theta} \qquad (4)$$

$$N_c = \frac{L_c}{d_{002}} \qquad (5)$$

Where $\lambda$, $FWHM$ is the X-ray wavelength which is 0.154056 nm and the FWHM of the peaks (002 and 100) respectively.

**Table 6.** Interlayer distance, in-plane, and out-of-plane crystallite sizes of GnPs-P, GnPs-COOH, GnPs-35COOH, GnPs-NH3, GnPs-O, GnPs-N, GnPs-A and GnPs-CF

| GnPs-Samples | $(2\theta)°$ (002) | FWHM (002) $(2\theta)°$ | $(2\theta)°$ (100) | FWHM (100) $(2\theta)°$ | $d_{002}$ (nm) | $L_c$ (nm) | $L_a$ (nm) | $N_c$ |
|---|---|---|---|---|---|---|---|---|
| GnPs-P | 26.50 | 0.4734 | 44.56 | 1.0986 | 0.3360 | 17.05 | 7.73 | 51 |
| GnPs-NH3 | 26.54 | 0.8367 | 44.64 | 3.2530 | 0.3356 | 9.65 | 2.61 | 29 |
| GnPs-A | 26.71 | 0.5326 | 44.63 | 4.0596 | 0.3350 | 15.15 | 2.09 | 45 |
| GnPs-COOH | 26.54 | 0.6214 | 44.63 | 4.3000 | 0.3356 | 13.04 | 1.98 | 39 |
| GnPs-35COOH | 26.36 | 2.5200 | 43.76 | 5.31 | 0.3378 | 3.200 | 1.59 | 9 |
| GnPs-CF | 26.54 | 0.6214 | 44.63 | 3.6255 | 0.3356 | 13.04 | 2.34 | 39 |
| GnPs-N | 26.37 | 0.5030 | 44.58 | 3.7426 | 0.3376 | 16.00 | 2.27 | 47 |
| GnPs-O | 26.36 | 0.6806 | 44.66 | 4.9943 | 0.3376 | 11.83 | 1.7 | 35 |

### 3.5. Molecular Dynamics (MD) Simulation of Pristine GnPs

### 3.6.1 Vibrational Properties of Pristine GnP (GnP-P)

This section reports the vibrational properties investigated from MD simulations carried out on trilayer graphene using the optimized Tersoff and Brenner empirical interatomic potential to describe the interactions between the carbon atoms[34]. The trilayer graphene model represents one platelet in our GnPs' aggregates (GnP-P), where one flake consists of a short stack of 3-6 layers of graphene. The theoretical lattice constant ($a^{theor.}$) of trilayer graphene was calculated to be 2.4856 Å using the optimized Tersoff and Brenner empirical potential which agrees well with the experimental lattice constant of graphene, 2.46 Å[31]. The lengths of the simulation box used in the x and y directions are $L_X$ = 24.85 Å, $L_Y$ = 21.52 Å, with a vacuum region of 35 Å applied in the z direction to avoid interaction between the periodic images. A time step of 0.002 ps was used during the micro canonical ensemble simulation (NVE) for proper equilibration of the system with a total run time of 16 ns. Periodic boundary conditions are employed in x, y, and z directions to eliminate boundary effects caused by finite system size.

GnP-P has six atoms in its primitive unit cell therefore resulting in a total number 18 phonon branches. There are a total of 15 optical phonon branches and 3 acoustic phonon branches. As shown in Fig. 10(A), the optical phonon branches (longitudinal optical (LO), transverse optical (TO) and out-of-plane optical (ZO)) are split into 3 optical phonon branches each whereas the acoustic phonon branches (longitudinal acoustic (LA), transverse acoustic (TA) and out-of-plane acoustic (ZA)) are split 2 optical phonon branches and 1 acoustic phonon branch each. The terms 'longitudinal and transverse' imply that the carbon atoms are displaced in directions parallel and perpendicular to the wavevector (q) of the phonons and in plane of the GnP-P platelet, respectively. The term "out-of-plane (Z)" implies the carbon atoms are displaced in directions perpendicular to the wavevector (q) of the phonons and out-of-plane of the GnP-P sheet.

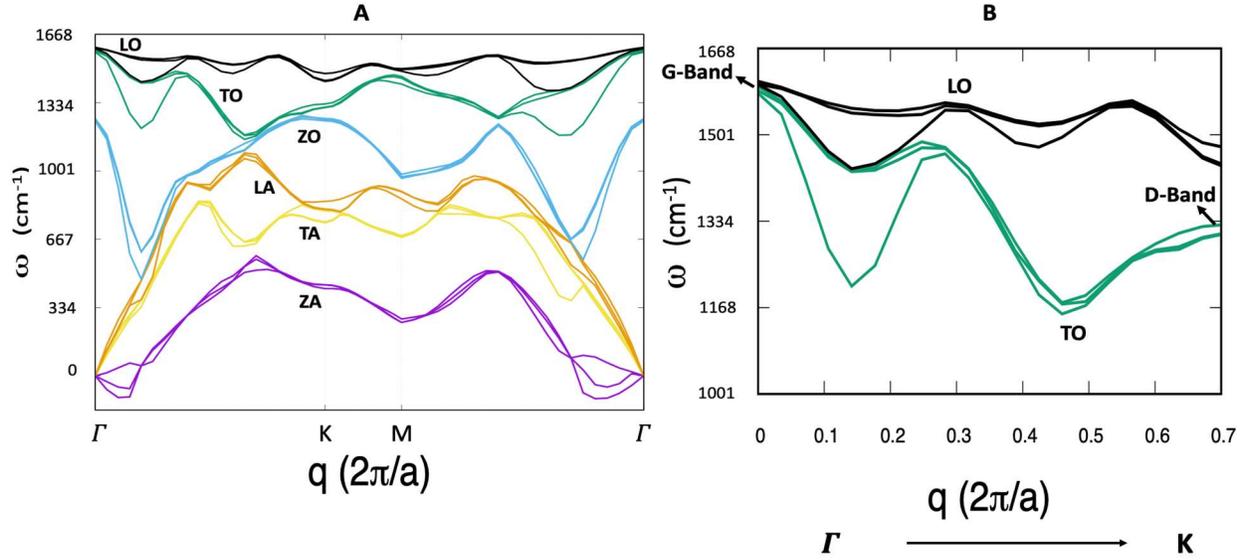

**Figure 10.** Calculated phonon dispersion relation of GnPs-P showing the (A) LO, TO, ZO, LA, TA and ZA phonon branches (B) LO and TO phonon branches only which contribute to D, G and 2D Raman peaks.

The phonon dispersion curves (LO, TO, ZO, LA, TA and ZA) of trilayer graphene are plotted with respect to the Brillouin zone points (namely Γ, M and K) and shown in Fig. 10(A). However, emphasis is placed on the in-plane optical vibrational modes at the Γ and K points (LO and TO) shown in Fig. 10(B) because they play a crucial role in studying the Raman and infrared (IR) spectra of GnPs-P. At the Γ-point, the optical modes are decomposed into $B_{2g}$, $E_{2g}$, $A_{2u}$ and $E_{1u}$ vibrational modes while the acoustic modes are decomposed into $E_{1u}$ and $A_{2u}$ modes. The $E_{2g}$ and $E_{1u}$ doubly degenerate modes are Raman and infrared active modes respectively. The $A_{2u}$ mode is an infrared active mode and the $B_{2g}$ mode is an optically inactive mode. The $E_{2g}$ mode at the Γ-point (TO + LO modes) and the $A_{1g}$ mode at the K-point (TO mode) are the Raman G and D peaks respectively shown in Fig. 11. The $E_{2g}$ mode at the Γ-point of GnPs-P evolves into: $E_{2g} = 2E_{2g} + E_{1u}$, while the $A_{1g}$ mode at the K-point evolves into: $A_{1g} = 2E + A_{1g}$[35] Table 7 shows the frequencies of the various modes and their values, around both the G and K points, that one could compare to the predicted values. The general agreement exhibited by the results support the notion that MD simulations can be used in analyzing the Raman spectra of GnPs-P.

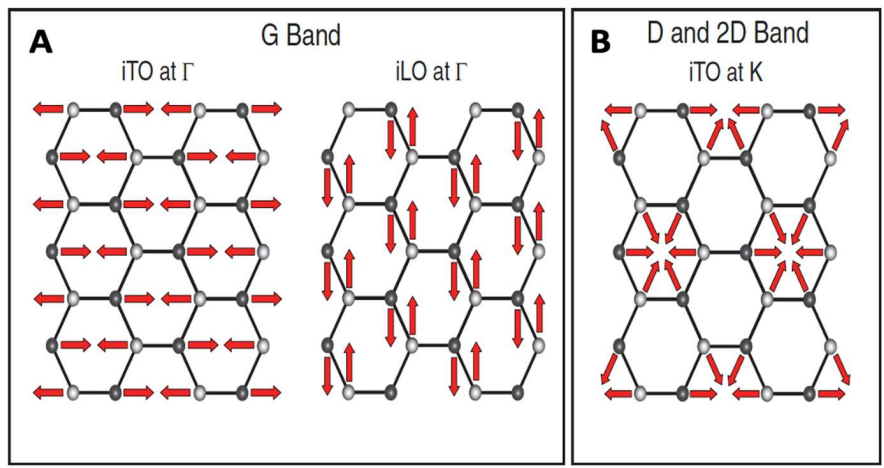

**Figure 11.** (A) G band vibration modes of the TO and LO phonon branches at the Γ-point (B) D band vibration mode of the TO phonon branch at the $K$-point[36].

**Table 7.** Calculated and experimental values of Raman and Infrared peaks of GnPs-P. R: Raman active; IR: Infrared active

|   | Vibrational Modes | GnP-P (Theoretical Values) | GnPs-P (Experimental Values) |
|---|---|---|---|
| Γ | $E_{2g}$ (R) | 1586 cm$^{-1}$ | 1581 cm$^{-1}$ (G Peak) |
|   | $E_{1u}$ (IR) | 1594 cm$^{-1}$ | 1670 cm$^{-1}$ (C=C bond from IR spectra) |
|   | $E_g$ (R & IR) | 1581 cm$^{-1}$ |  |
| K | $A_{1g}$ (R) | 1328 cm$^{-1}$ (D peak) 2656 cm$^{-1}$ (2D peak) | 1353 cm$^{-1}$ (D peak) 2722 cm$^{-1}$ (2D peak) |
|   | E (R & IR) | 1310 cm$^{-1}$ |  |
|   | E (R & IR) | 1312 cm$^{-1}$ |  |

### 3.6.2 Adsorption Properties of Pristine GnPs

The adsorption properties of NO$_2$ on GnP-P (consisting of three graphene layers) was investigated using MD simulations within the LAMMPS software code. A simulation space was given dimensions of 34 Å × 34 Å × 60 Å in the X, Y, and Z directions, respectively. The GnP-P was placed at the bottom of the box as shown in Fig. 12 and the boundary conditions were set to be periodic in all directions. Fifty NO$_2$ molecules were randomly deposited inside the simulation space and energy minimization was performed to relax the NO₂ molecules. Equilibration of the NO₂ molecules only was then carried out using the canonical ensemble (NVT) to uniformly distribute the molecules for 0.1 ns. After this step, interactions between the NO₂ molecules and GnP-P were turned on and the total system was equilibrated for 0.5 s to adsorb the molecules. The number of adsorbed molecules were obtained by counting the number of molecules that are in a region of height 5Å above GnP-P. The Lorentz-Bertholet combination rules in equations [6] and [7] was used to calculate the Lennard Jones potential parameters describing interactions between GnP-P and NO₂ (see table 8).

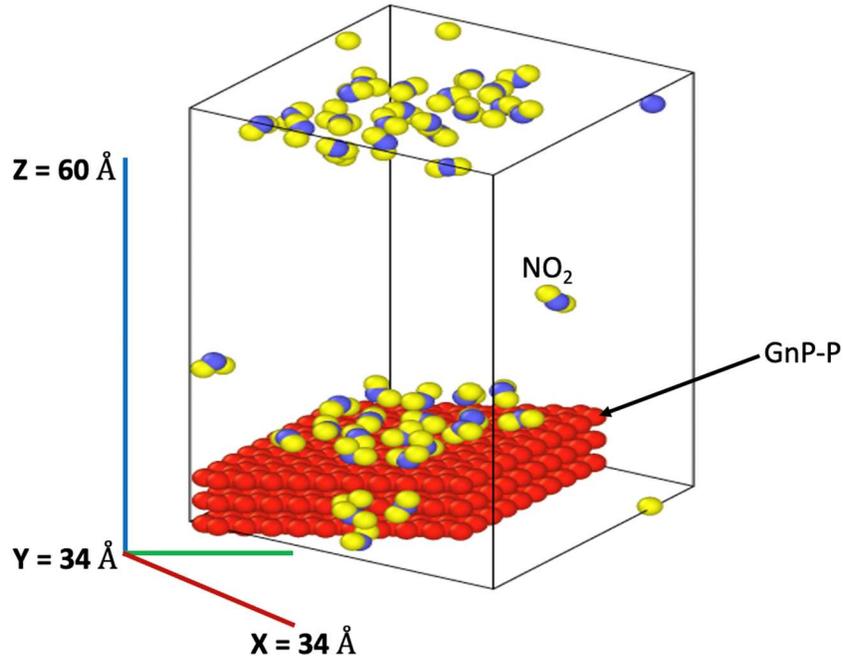

**Figure 12.** NO₂ molecules adsorbed on GnP-P consisting of three layers of graphene

$$\varepsilon_{ij} = \sqrt{\varepsilon_{ii}\varepsilon_{jj}} \tag{6}$$

$$\sigma_{ij} = \frac{1}{2}(\sigma_{ii} + \sigma_{jj}) \tag{7}$$

Where $\varepsilon_{ij}$ and $\sigma_{ij}$ are the interaction energy and distance between particles i and j, respectively.

**Table 8.** Lennard Jones Parameters for NO₂-GnP-P interactions

| Adsorbate | ε (kcal/mol) | σ (Å) |
|---|---|---|
| C-N | 0.14 | 3.10 |
| N-N | 0.16 | 3.12 |
| O-O | 0.20 | 2.85 |
| N-O | 0.18 | 2.99 |
| C-O | 0.16 | 2.90 |

The adsorption energy $E_a$ is defined in equation [8]:

$$E_a = E_{(NO2 + GnP-P)} - (E_{(GnP-P)} + E_{(NO2)}) \qquad (6)$$

Where $E_{(NO2 + GnP-P)}$ is the total energy of the graphene adsorbed system, $E_{(GnP-P)}$ is the energy of GnP-P, and $E_{(NO2)}$ is the energy of the $NO_2$ molecule. The adsorption energy of $NO_2$ molecule on GnP-P was found to be -97 kcal/mol, with the negative value implying that adsorption has taken place and was an exothermic reaction. Also, our results show that 28 $NO_2$ molecules were adsorbed on the GnP-P sheet.

## 4. CONCLUSION

In this paper, the structural and surface composition properties of 8 species of GnPs were investigated comprehensively via SEM, Raman spectroscopy, XRD and XPS. Also, MD simulations were performed on a single platelet sheet consisting of three layers of graphene to calculate the Raman and infrared vibrational frequencies and the adsorption energy. Functionalizing GnPs with the aforementioned functional groups using the dielectric barrier discharge plasma method reduces the particle and crystallite sizes of the GnPs samples as shown in our SEM and XRD results.

The GnPs-35COOH have delaminated edges and a highly disordered structure due to the high concentration of oxygen containing functional groups present. Raman data showed no shift in the Raman G peak of the functionalized GnPs, implying that the functional groups were not causing strain within the structure. The sharp XRD peaks of the low-density functionalized samples suggested that the crystallinity of the samples was retained but at a higher concentration and that the structural quality was lowered and highly disordered. In addition, introducing higher concentrations of COOH (35 wt.%) to GnPs-P introduces *n*-type doping which is evident in the broadening and a redshift of the 2D peak. The XPS measurements confirmed the existence of a low concentration of the functional groups present in the low-density functionalized samples. MD simulations were utilized to evaluate the Raman and IR active modes present in GnPs-P and agreed well with our experimental results. Lastly, the adsorption properties of $NO_2$ on GnPs-P was determined using MD simulations. These results will help contribute to the design and production of the GnP-P-based sensor for $NO_2$ gas detection.

AUTHOR INFORMATION

**Corresponding Author**

* Correspondence to: pmisra@howard.edu

**Author Contributions**

H. A. supported with performing Raman measurements, M. L. K. and A. J. B. performed x-ray photoelectron spectroscopy measurements, A. V. D. performed x-ray diffraction measurements on GnPs-35COOH sample, A. F. R. and P. M. assisted with the analyses, support, and general project oversight.

REFERENCES

1. M Mittal and A Kumar, "Carbon nanotube (CNT) gas sensors for emissions from fossil fuel burning," Sensors and Actuators B: Chemical **203**, 349-362 (2014).
2. Tao Han, Anindya Nag, Subhas Chandra Mukhopadhyay, and Yongzhao Xu, "Carbon nanotubes and its gas-sensing applications: A review," Sensors and Actuators A: Physical **291**, 107-143 (2019).
3. Wenchao Tian, Xiaohan Liu, and Wenbo Yu, "Research progress of gas sensor based on graphene and its derivatives: a review," Applied Sciences **8** (7), 1118 (2018).
4. Shengxue Yang, Chengbao Jiang, and Su-huai Wei, "Gas sensing in 2D materials," Applied Physics Reviews **4** (2), 021304 (2017).
5. K. Kalaitzidou, H. Fukushima, L.T. Drzal, "Multifunctional polypropylene composites produced by incorporation of exfoliated graphite nanoplatelets" Carbon N Y, **45** (2007), pp. 1446-1452.
6. S. Kim, I. Do, L.T. Drzal, "Multifunctional xGnP/LLDPE nanocomposites prepared by solution compounding using various screw rotating systems," Macromol. Mater. Eng., **294** (2009), pp. 196-205.
7. Rigosi AF, Kruskopf M, Hill HM, Jin H, Wu BY, Johnson PE, Zhang S, Berilla M, Walker AR, Hacker CA, Newell DB. Gateless and reversible Carrier density tunability in epitaxial graphene devices functionalized with chromium tricarbonyl. Carbon. 2019 Feb 1;142:468-74.
8. Rigosi AF, Patel D, Marzano M, Kruskopf M, Hill HM, Jin H, Hu J, Walker AR, Ortolano M, Callegaro L, Liang CT. Atypical quantized resistances in millimeter-scale epitaxial graphene pn junctions. Carbon. 2019 Dec 1;154:230-7.
9. Rigosi AF, Elmquist RE. The quantum Hall effect in the era of the new SI. Semiconductor science and technology. 2019 Aug 27;34(9):093004.
10. Andrea Ponzoni, Camilla Baratto, Nicola Cattabiani, Matteo Falasconi, Vardan Galstyan, Estefania Nunez-Carmona, Federica Rigoni, Veronica Sberveglieri, Giulia Zambotti, and Dario Zappa, "Metal oxide gas sensors, a survey of selectivity issues addressed at the SENSOR Lab, Brescia (Italy)," Sensors **17** (4), 714 (2017).
11. Zuquan Wu, Xiangdong Chen, Shibu Zhu, Zuowan Zhou, Yao Yao, Wei Quan, and Bin Liu, "Enhanced sensitivity of ammonia sensor using graphene/polyaniline nanocomposite," Sensors and Actuators B: Chemical **178**, 485-493 (2013).
12. Filippo Pinelli, Tommaso Nespoli, Andrea Fiorati, Silvia Farè, Luca Magagnin, and Filippo Rossi, "Graphene nanoplatelets can improve the performances of graphene oxide–polyaniline composite gas sensing aerogels," Carbon Trends, 100123 (2021).
13. Run Zhang, Jian-Bo Jia, Jian-Liang Cao, and Yan Wang, "SnO2/Graphene Nanoplatelet Nanocomposites: Solid-State Method Synthesis With High Ethanol Gas-Sensing Performance," Frontiers in chemistry **6**, 337 (2018).
14. Ling Ti Kong, "Phonon dispersion measured directly from molecular dynamics simulations," Computer Physics Communications **182** (10), 2201-2207 (2011).
15. Large-scale Atomic and Molecular Massively Parallel Simulator, "Lammps," available at: http:/lammps. sandia. gov (2013).
16. Steve Plimpton, "Fast parallel algorithms for short-range molecular dynamics," Journal of computational physics **117** (1), 1-19 (1995).

17  David Nečas and Petr Klapetek, "Gwyddion: an open-source software for SPM data analysis," Open Physics **10** (1), 181-188 (2012).
18  Danijela V Brković, Vesna V Kovačević, Goran B Sretenović, Milorad M Kuraica, Nemanja P Trišović, Luca Valentini, Aleksandar D Marinković, José M Kenny, and Petar S Uskoković, "Effects of dielectric barrier discharge in air on morphological and electrical properties of graphene nanoplatelets and multi-walled carbon nanotubes," Journal of Physics and Chemistry of Solids **75** (7), 858-868 (2014).
19  Seyyedeh Saadat Shojaeenezhad, Mansoor Farbod, and Iraj Kazeminezhad, "Effects of initial graphite particle size and shape on oxidation time in graphene oxide prepared by Hummers' method," Journal of Science: Advanced Materials and Devices **2** (4), 470-475 (2017).
20  In-Yup Jeon, Hyun-Jung Choi, Sun-Min Jung, Jeong-Min Seo, Min-Jung Kim, Liming Dai, and Jong-Beom Baek, "Large-scale production of edge-selectively functionalized graphene nanoplatelets via ball milling and their use as metal-free electrocatalysts for oxygen reduction reaction," Journal of the American Chemical Society **135** (4), 1386-1393 (2013).
21  D Sfyris, GI Sfyris, and C Galiotis, "Stress intrepretation of graphene E-2g and A-1g vibrational modes: theoretical analysis," arXiv preprint arXiv:1706.04465 (2017).
22  Nonjabulo PD Ngidi, Moses A Ollengo, and Vincent O Nyamori, "Effect of doping temperatures and nitrogen precursors on the physicochemical, optical, and electrical conductivity properties of nitrogen-doped reduced graphene oxide," Materials **12** (20), 3376 (2019).
23  L Gustavo Cançado, A Jorio, EH Martins Ferreira, F Stavale, Carlos Alberto Achete, Rodrigo Barbosa Capaz, Marcus Vinicius de Oliveira Moutinho, Antonio Lombardo, TS Kulmala, and Andrea Carlo Ferrari, "Quantifying defects in graphene via Raman spectroscopy at different excitation energies," Nano letters **11** (8), 3190-3196 (2011).
24  Jiang-Bin Wu, Miao-Ling Lin, Xin Cong, He-Nan Liu, and Ping-Heng Tan, "Raman spectroscopy of graphene-based materials and its applications in related devices," Chemical Society Reviews **47** (5), 1822-1873 (2018).
25  A Das, S Pisana, S Piscanec, B Chakraborty, SK Saha, UV Waghmare, R Yiang, HR Krishnamurhthy, AK Geim, and AC Ferrari, "Electrochemically top gated graphene: Monitoring dopants by Raman scattering," arXiv preprint arXiv:0709.1174 (2007).
26  C Casiraghi, "Doping dependence of the Raman peaks intensity of graphene close to the Dirac point," Physical Review B **80** (23), 233407 (2009).
27  José M Caridad, Francesco Rossella, Vittorio Bellani, Marco Maicas, Maddalena Patrini, and Enrique Díez, "Effects of particle contamination and substrate interaction on the Raman response of unintentionally doped graphene," Journal of Applied Physics **108** (8), 084321 (2010).
28  CRSV Boas, B Focassio, E Marinho, DG Larrude, MC Salvadori, C Rocha Leão, and Demetrio J Dos Santos, "Characterization of nitrogen doped graphene bilayers synthesized by fast, low temperature microwave plasma-enhanced chemical vapour deposition," Scientific reports **9** (1), 1-12 (2019).
29  Cui Zhang, Daniel M Dabbs, Li-Min Liu, Ilhan A Aksay, Roberto Car, and Annabella Selloni, "Combined effects of functional groups, lattice defects, and edges in the infrared spectra of graphene oxide," The Journal of Physical Chemistry C **119** (32), 18167-18176 (2015).

30  Karim Kakaei, Mehdi D Esrafili, and Ali Ehsani, *Graphene surfaces: particles and catalysts*. (Academic Press, 2018).
31  M Mohr, J Maultzsch, E Dobardžić, S Reich, I Milošević, M Damnjanović, A Bosak, M Krisch, and C Thomsen, "Phonon dispersion of graphite by inelastic x-ray scattering," Physical Review B **76** (3), 035439 (2007).
32  Adriyan Milev, Michael Wilson, GS Kamali Kannangara, and Nguyen Tran, "X-ray diffraction line profile analysis of nanocrystalline graphite," Materials Chemistry and Physics **111** (2-3), 346-350 (2008).
33  It-Meng Low, Hani Manssor Albetran, and Michael Degiorgio, "Structural characterization of commercial graphite and graphene materials," Journal of Nanotechnology and Nanomaterials **1** (1) (2020).
34  L Lindsay and DA Broido, "Optimized Tersoff and Brenner empirical potential parameters for lattice dynamics and phonon thermal transport in carbon nanotubes and graphene," Physical Review B **81** (20), 205441 (2010).
35  Jia-An Yan, WY Ruan, and MY Chou, "Phonon dispersions and vibrational properties of monolayer, bilayer, and trilayer graphene: Density-functional perturbation theory," Physical review B **77** (12), 125401 (2008).
36  Yucheng Lan, Mobolaji Zondode, Hua Deng, Jia-An Yan, Marieme Ndaw, Abdellah Lisfi, Chundong Wang, and Yong-Le Pan, "Basic concepts and recent advances of crystallographic orientation determination of graphene by Raman spectroscopy," Crystals **8** (10), 375 (2018).